\documentclass[prl,twoside,twocolumn,aps,amsmath,amssymb]{revtex4}

\usepackage{bbm,mathrsfs}
\usepackage{graphicx}
\usepackage{amsfonts}
\usepackage{amsthm}

\def\textbf#1{{\bf #1}}
\def\be{\begin{equation}}
\def\ee{\end{equation}}
\def\ben{\begin{eqnarray}}
\def\een{\end{eqnarray}}
\def\eea{\end{array}}
\def\bea{\begin{array}}
\newcommand{\Tr}[0]{\mathrm{Tr}}
\newcommand{\ot}[0]{\otimes}
\newcommand{\bei}{\begin{itemize}}
\newcommand{\eei}{\end{itemize}}
\newcommand{\ket}[1]{|#1\rangle}

\def\blacksquare{\vrule height 4pt width 3pt depth2pt}

\theoremstyle{definition}

\newcommand{\ie}{{\it{i.e.}}}

\begin{document}

\title{Bell inequalities with no quantum violation and unextendible product bases}

\author{R. Augusiak$^{1}$, J. Stasi\'nska$^{2}$, C. Hadley$^{1}$, J. K. Korbicz$^{1}$,
M. Lewenstein$^{1,3}$ and A. Ac\'in$^{1,3}$}
\affiliation{$^{1}$ICFO--Institut de Ci\`{e}ncies Fot\`{o}niques,
08860 Castelldefels (Barcelona), Spain} \affiliation{$^{2}$Grup de
F\'isica Te\`orica: Informaci\'o i Fen\`omens Qu\`antics,
Universitat Aut\`onoma de Barcelona, 08193 Bellaterra (Barcelona),
Spain} \affiliation{$^{3}$ICREA--Instituci\'o Catalana de Recerca
i Estudis Avan\c{c}ats, Lluis Companys 23, 08010 Barcelona, Spain}

\begin{abstract}
The strength of classical correlations is subject to certain
constraints, commonly known as Bell inequalities. Violation of
these inequalities is the manifestation of
non-locality---displayed, in particular, by quantum mechanics,
meaning that quantum mechanics can outperform classical physics at
tasks associated with such Bell inequalities. Interestingly,
however, there exist situations in which this is not the case. We
associate an intriguing class of bound entangled states,
constructed from unextendable product bases (UPBs) with a wide
family of tasks, for which (i) quantum correlations do not
outperform the classical ones but (ii) there exist supraquantum
nonsignalling correlations that do provide an advantage.
\end{abstract}

 \pacs{}

\maketitle {\it Introduction.---}The existence of correlations is
an inherent property of composite physical systems and, as such,
is fundamental for our understanding of physical phenomena. On the
other hand, physical principles impose limits on the correlations
between the results of measurements performed on distant systems.
If the measurements correspond to spacelike separated events, the
observed correlations should obey the principle of no signaling,
which prevents any faster-than-light communication among the
parties. If the systems are quantum, it should be possible to
write the correlations as results of local measurements acting on
a global quantum state. Finally, the observed correlations are
said to be classical if they are attainable with shared classical
randomness. All three kinds of correlations correspond to sets of
probabilities of measurement outcomes and, as such, form convex
sets, represented schematically in Fig. 1.

Bell was the first to point out that classical correlations (CC)
are constrained by certain inequalities (the famous Bell
inequalities) \cite{Bell}. Correlations which violate a Bell
inequality, and thus do not correspond to any classical model, are
known as nonlocal. Bell's theorem guarantees the existence of
quantum correlations (QC) that are nonlocal. However, it is known
also that there are nonsignaling correlations (NC) which are
supraquantum~\cite{PRbox}, i.e., not attainable by measurements
acting on a quantum state, yet violating Bell inequalities.

Apart from its fundamental importance, understanding the relation
among the sets of correlations is crucial from a practical point
of view, since correlations and application as an information
resource. In particular, one of the goals of quantum information
theory is to understand when QC give an advantage over CC. For
instance, nonlocal QC provide cryptographic security not
achievable\linebreak with classical theory
\cite{Ekert91PRL,AcinDevice}. They can also be used to certify the
presence of randomness~\cite{randomness} and outperform CC at
communication complexity problems \cite{B10RMP}.

\begin{figure}[t]
a)\includegraphics[width=0.22\textwidth]{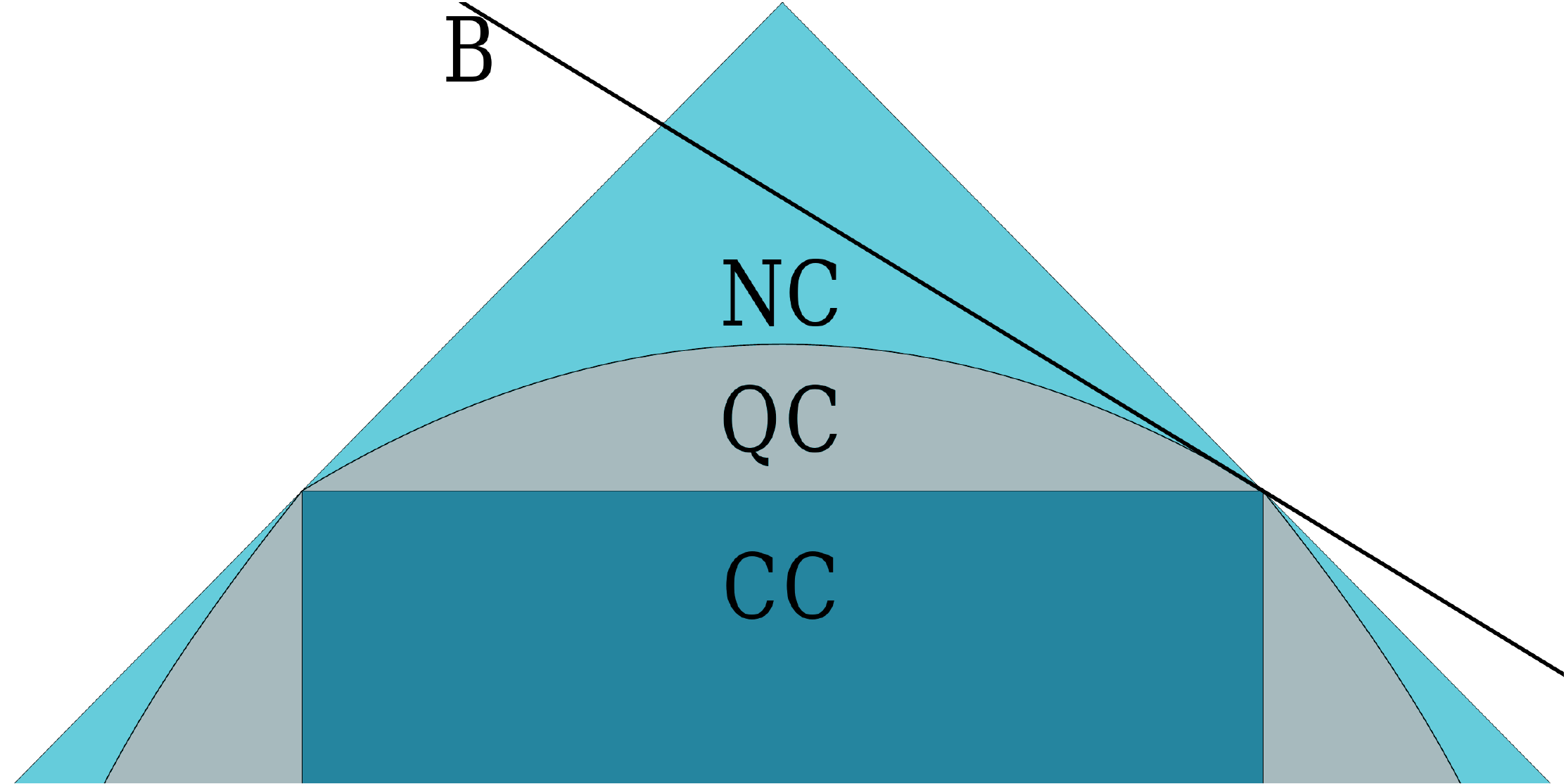}\,\,b)\includegraphics[width=0.22\textwidth]{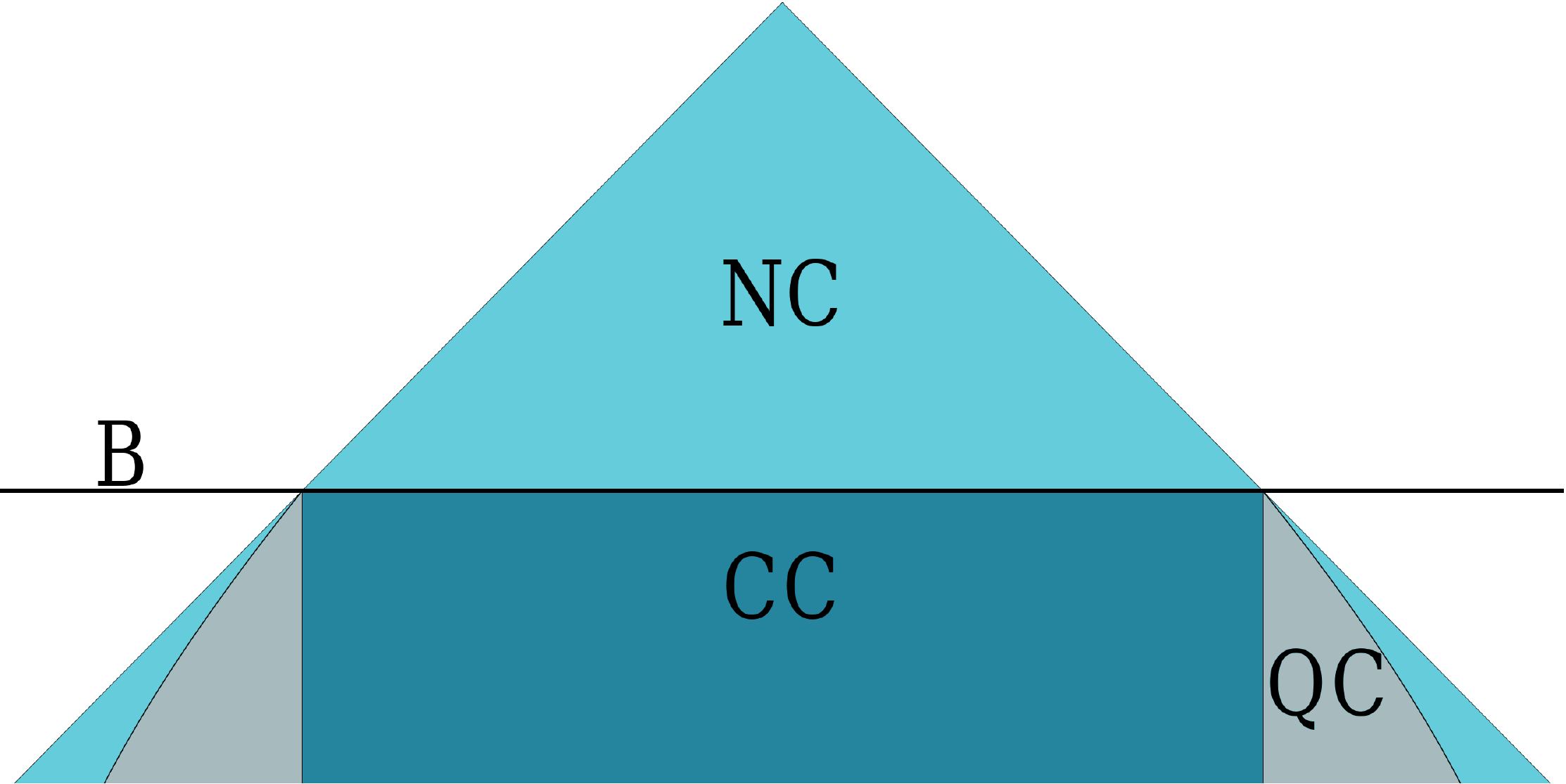}
\caption{Schematic depiction of the sets of classical ($CC$),
quantum ($QC$) and non-signalling correlations ($NC$). Tight Bell
inequalities correspond to facets of the classical set. $B$
denotes a Bell inequality with no quantum violation which is a)
not tight, b) tight. Note that a tight Bell inequality guarantees
that a region in which quantum and classical correlations coincide
is a facet. A non-tight inequality may define a common region of
the classical and quantum sets that does not have maximal
dimension.}\label{fig1}
\end{figure}
While QC are in general more powerful than CC, there are some
intriguing situations in which CC and QC perform equally well.
This equivalence can be detected by Bell inequalities which are
not violated by QC. The first examples of such inequalities were
given in Ref.~\cite{Linden07PRL} for two parties. Unfortunately,
none of these is tight~\cite{prcomm}. The importance of tight Bell
inequalities stems from the fact that they correspond to facets of
the convex set (polytope) of classical correlations (see Fig.
\ref{fig1}) and hence are sufficient to fully characterize it
\cite{conv_poli}. Multipartite Bell inequalities with no quantum
violation were later provided in Ref.~\cite{Almeida10PRL} and,
moreover, shown to be tight for $3\leq n\leq7$ parties. Apart from
these examples, we know little about information tasks, or
equivalently Bell inequalities, where QC do not provide any
advantage.

In this Letter, we demonstrate an {\it a priori} unexpected
relation between such inequalities and unextendable product bases
(UPBs) \cite{BennettUPBPRL}. Recall that the latter is a
collection of orthogonal product vectors spanning a proper
subspace $V$ of some $n$-partite Hilbert space $\mathcal{H}$, such
that there does not exist any other product vector in
$\mathcal{H}$ orthogonal to $V$. The fundamental physical
importance of UPBs stems from the fact that they allow for the
construction \cite{BennettUPBPRL} of one of the first examples of
bound (i.e., nondistillable) entangled states—--being one of the
most intriguing concepts in quantum information theory
\cite{Horodeccy}--—and, furthermore, that they give rise to {\it
nonlocality without entanglement} \cite{BennettPRA}, i.e., the
impossibility of perfect distinguishability of some orthogonal
product states by means of local operations and classical
communication.

Below, we prove how UPBs satisfying a given requirement give rise
to Bell inequalities without a quantum violation. Our construction
covers some of the inequalities previously derived in
Ref.~\cite{Almeida10PRL} and, thus, can lead to tight Bell
inequalities. Moreover, the construction can be exploited in the
opposite direction: We derive new examples of UPBs from some of
the Bell inequalities with no quantum violation from
Ref.~\cite{Almeida10PRL}.

The considered scenario consists of $n$ distant observers having
access to $n$ systems. Observer $i$ $(i=1,\ldots,n$) can perform
on his system one of $m_i$ possible measurements. His choice of
measurement is denoted $x_i=1,\ldots,m_i$, and the obtained result
$a_i=1,\ldots,r_i^{x_{i}}$ with $r_i^{x_{i}}$ denoting the number
of outcomes the $m_i$th measurement has, while
$\textbf{x}=(x_1,\ldots,x_n)$ and $\textbf{a}=(a_1,\ldots,a_n)$
stand for the corresponding vectors. The correlations among the
parties are described by the conditional probability
$p(\textbf{a}|\textbf{x})$.

Consider a linear combination of these probabilities defined by a
$2n$-index tensor $T_{\textbf{x,a}}$, $\sum
T_{\textbf{x,a}}p(\textbf{a}|\textbf{x})$. It leads to the Bell
inequality
$\sum_{\textbf{x,a}}T_{\textbf{x,a}}p(\textbf{a}|\textbf{x})\leq
\beta_{\mathrm{C}}$, where $\beta_{\mathrm{C}}$
($\beta_{\mathrm{Q}}$, $\beta_{\mathrm{N}}$) is its maximal value
for CC (QC, NC). It is non-trivial if it is not an inequality for
general non-signalling correlations, \ie,
$\beta_{\mathrm{N}}>\beta_{\mathrm{C}}$, while it is violated by
QC if $\beta_{\mathrm{Q}}>\beta_{\mathrm{C}}$. Any Bell inequality
can also be seen as a non-local game, in which the parties are
given the input $\textbf{x}$ and have to produce the output
$\textbf{a}$, in a distributed manner, all the possibilities being
weighted by the tensor $T_{\textbf{x,a}}$. The maximum values of
the inequalities give the optimal winning probability for the
different sets of correlations.

{\it From a UPB to Bell inequalities.---}Consider an $n$-partite product
Hilbert space $\mathcal{H}=\ot_{i=1}^{n}\mathbbm{C}^{d_i}$ and a set
$\mathcal{S}$ of orthogonal product vectors from $\mathcal{H}$:
$\mathcal{S}=\{\ket{\psi_j^{(1)}}\otimes\ldots\otimes\ket{\psi_j^{(n)}}\}_{j=1}^{|\mathcal{S}|},$
with $\ket{\psi_j^{(i)}}\in\mathbbm{C}^{d_i}$. The $|\mathcal{S}|$ local states
for each party constitute $n$ sets, denoted by $\mathcal{S}^{(i)}$.

We partition each $\mathcal{S}^{(i)}$ into disjoint subsets
$\mathcal{S}_k^{(i)}$ such that all vectors forming a particular
subset are mutually orthogonal. Each $\mathcal{S}_k^{(i)}$ defines
a measurement, while the different vectors within every such
subset are associated to the measurement outcomes. In order to
remove the ambiguity in splitting the local sets
$\mathcal{S}^{(i)}$ into subsets, we restrict ourselves to sets
$\mathcal{S}$ having the property that {\it no two vectors
belonging to different subsets $\mathcal{S}^{(i)}_k$ are
orthogonal.} Below we refer to this property as (P). This
constraint is automatically satisfied in the case of qubits.

It is straightforward to assign to each vector from $\mathcal{S}$ a conditional
probability $p(\boldsymbol{a}_j|\boldsymbol{x}_j)$: the measurement by the
observer $i$ is given by the index $k$, enumerating the subset
$\mathcal{S}^{(i)}_k$ to which $\ket{\psi_j^{(i)}}$ belongs, while the result
corresponds to the position of this state within the set. We now consider
linear combinations of these conditional probabilities with weights $q_j$. The
maximum of such a linear combination over all local strategies is
$\beta_{\mathrm{C}}=\max\{q_j\}$. Indeed, due to (P), orthogonality of any two
vectors from $\mathcal{S}$ means that at some position they have different
vectors from the same local subset. At the level of probabilities, this means
that if one of them, say $p(\boldsymbol{a}_j|\boldsymbol{x}_j)$, equals unity,
the rest have to be zero, as they always have at some position the same input
but a different output. Then, we get the Bell inequality
\begin{equation}\label{genBell}
\sum_{j}q_{j}p(\boldsymbol{a}_{j}|\boldsymbol{x}_{j})\leq \max\{q_j\},
\end{equation}
from the initial set of orthogonal product vectors $\mathcal{S}$.

It is now easy to prove that all these inequalities are not violated by QC.

\noindent{\bf Fact 1.} {\it Let $\mathcal{S}$ be a set of
orthogonal product vectors possessing the property (P). Then for
the corresponding Bell inequality (\ref{genBell}) it holds that
$\beta_{\mathrm{C}}=\beta_{\mathrm{Q}}=\max\{q_i\}$.}

\noindent{\bf Proof.} First of all, since the dimension is arbitrary, we
can restrict the analysis to projective measurements. Let us assign projectors
$P_{j}^{(i)}$ (we enumerate them in the same way as the vectors
$\ket{\psi_j^{(i)}}$) to the outcomes of the local observables, and construct
the Bell operator
\begin{equation}\label{BellOp}
B=\sum_{j=1}^{|\mathcal{S}|}q_j\bigotimes_{i=1}^{n} P_{j}^{(i)}.
\end{equation}
In general, $P_{j}^{(i)}$ may be different from the local vectors
of $\mathcal{S}$ and, moreover, they can be degenerate.
Nevertheless, they maintain the orthogonality of the local vectors
from $\mathcal{S}$. Precisely, any pair of projectors
$\otimes_{i=1}^{n}P_j^{(i)}$ have at some position local
projectors corresponding to the same observable but different
outcomes, meaning that all of them are orthogonal. Thus the
maximal eigenvalue of $B$ is $\max\{q_i\}$, and hence
$\beta_{\mathrm{Q}}=\max\{q_i\}$. $\blacksquare$

Our construction offers a systematic and easy way of generating
Bell inequalities with no quantum violation from orthogonal
product vectors. However, it could be the case that all the
derived inequalities are trivial, in the sense of not being
violated by any NC. Here is where the concept of a UPB becomes
relevant.

\noindent{\bf Fact 2.} {\it If $\mathcal{S}$ is a UPB with the property (P),
the resulting Bell inequality (\ref{genBell}) with $q_j=1$ is violated by NC.}

\noindent{\bf Proof.} Let $\Pi_{\mathrm{UPB}}$ be a projector onto
the subspace of $\mathcal{H}$ spanned by $\mathcal{S}$. Then, the
Bell operator $B$ with $q_j=1$ and the measurements defined by
$\mathcal{S}$ is exactly $\Pi_{\mathrm{UPB}}$.

Consider now the normalized entanglement witness
$W=[1/(|\mathcal{S}|-\epsilon
D)]\left(\Pi_{\mathrm{UPB}}-\epsilon\mathbbm{1}\right)$ with
$D=\mathrm{dim}\mathcal{H}$ and
$\epsilon=\min_{\psi_{\mathrm{prod}}}\langle\psi_{\mathrm{prod}}|\Pi_{\mathrm{UPB}}|\psi_{\mathrm{prod}}\rangle$.
This witness detects the bound entangled state
$\varrho=(\mathbbm{1}-\Pi_{\mathrm{UPB}})/(D-|\mathcal{S}|)$. A
direct check shows that
$\Tr(BW)=\Tr(\Pi_{\mathrm{UPB}}W)=|\mathcal{S}|(1-\epsilon)/(|\mathcal{S}|-\epsilon
D)$, which is greater than one (in this case
$\beta_{C}=\beta_{Q}=1$) whenever $|\mathcal{S}|>\epsilon D$. The
latter, however, follows from the very definition of $W$.
Consequently, $W$ violates the Bell inequality resulting from
$\mathcal{S}$. This completes the proof, as local measurements
acting on a witness give raise to NC (see, e.g., Refs.
\cite{Barnum10PRL,Hadley10PRL}). $\blacksquare$


Our construction, then, shows how to derive non-trivial Bell
inequalities with no quantum violation from any UPB with property
(P). As mentioned, this property is always satisfied in the case
of qubits. For two parties there exists no qubit UPB. Moving to
three parties, it was shown in Ref.~\cite{Bravyi04QIP}, that by
local unitaries and permutations of particles, all UPBs can be
brought to:
$\mathcal{S}=\{\ket{000},\ket{1e_2e_3},\ket{e_11e_3^{\perp}},\ket{e_1^{\perp}e_2^{\perp}1}\}$
with $\ket{e_i}\neq \ket{0},\ket{1}$ and $\langle
e_i^{\perp}|e_i\rangle=0$ ($i=1,2,3$). Following the above rules,
at each site we can define two subsets of mutually orthogonal
vectors, namely, $\mathcal{S}_0=\{\ket{0},\ket{1}\}$ and
$\mathcal{S}_1^{(i)}=\{\ket{e_i},\ket{e_i^{\perp}}\}$. Then, we
assign to each element in the UPB the following probabilities:
$\ket{000}\rightarrow p(000|000)$, $\ket{1e_2e_3}\rightarrow
p(100|011)$, $\ket{e_11e_3^{\perp}}\rightarrow p(011|101)$, and
$\ket{e_1^{\perp}e_2^{\perp}1}\rightarrow p(111|110)$. Adding
them, we obtain the inequality
$p(000|000)+p(100|011)+p(011|101)+p(111|110)\leq 1$ with
$\beta_\mathrm{Q}=\beta_\mathrm{C}=1$. This inequality, previously
derived in Ref.~\cite{Sliwa03PLA}, is one of the tight
inequalities studied in Ref. \cite{Almeida10PRL}. This shows that
our construction can lead to \emph{tight} Bell inequalities with
no quantum violation. In Ref.~\cite{moreUPB}, the above UPB was
generalized to an arbitrary odd number of qubits. We have checked
that the corresponding inequality for $n=5$, which is not the same
as the five-party Bell inequality of \cite{Almeida10PRL}, is not
tight.

Moving to dimensions larger than two, there already exist UPBs for
two parties. Although, as explained later, none of them has
property (P), there do exist examples for more than two parties
with this property, such as the UPB of Ref.~\cite{cerfniset}. We
applied our construction to these states in the case of four
three-dimensional system. Unfortunately, the resulting inequality
is not tight.

{\it From a Bell inequality to UPB.---}Clearly, the above
procedure can be applied in reverse: given a Bell
inequality~\eqref{genBell}, one can derive, following
analogous rules, a set of product vectors. The number of different
inputs at each position gives the number of different local
subsets, while the number of different outputs corresponding to a
particular input gives the number of elements of the corresponding
subset. The maximal number of different outputs at the $i$th
position gives the dimension of the local Hilbert space $d_i$.
Note that in the general case, the derived vectors are not
necessarily orthogonal. In what follows, we consider the set of
Bell inequalities with no quantum violation given
in~\cite{Almeida10PRL}. These inequalities are such that the
derived product vectors are orthogonal and naturally possesses the
property (P). Remarkably, as we will see shortly, they define a
new class of UPBs.

The explicit form of these inequalities for odd $n$ reads
\begin{equation}\label{BellIneqOdd}
\sum_{k=0}^{(n-1)/2} \sum_{i_1<\ldots<i_{2k}=1}^{n}D_{i_1\ldots
i_{2k}}p(\boldsymbol{0}|\boldsymbol{0})\leq 1,
\end{equation}
while for even $n$,
\begin{eqnarray}\label{BellIneqEven}
\sum_{k=0}^{(n-2)/2} \sum_{i_1<\ldots<i_{2k}=2}^{n}\hspace{-0.3cm}D_{i_1\ldots
i_{2k}}[p(\boldsymbol{0}|\boldsymbol{0})+p(0\ldots 01|10\ldots 0)]\leq 1.
\end{eqnarray}
Here $\boldsymbol{0}=(0,\ldots,0)$ and $D_{i_1,\ldots,i_k}$ flips
($0\leftrightarrow 1$) inputs and outputs at positions $i_1,\ldots,i_k$
and $i_1-1,\ldots,i_k-1$ (if $i_j=1$ then
$i_j-1=n$), respectively.

We now derive the product vectors corresponding to these
inequalities for arbitrary $n$. Note that all terms in Eqs.
(\ref{BellIneqOdd}) and (\ref{BellIneqEven}) have at each position
two possible incomes and outcomes. Thus, at each site we can
define a pair of two-element sets and, without any loss of
generality, we can take them to be equal for all sites; say
$\mathcal{S}_0=\{\ket{0},\ket{1}\}$ and
$\mathcal{S}_1=\{\ket{e},\ket{e^{\perp}}\}$ with $\ket{e}\neq
\ket{0},\ket{1}$.

Let $V$ denote a unitary operation such that $V\ket{0}=\ket{e}$
and $V\ket{1}=\ket{e^{\bot}}$, while $\sigma_x$, the standard
Pauli matrix flipping $\ket{0}\leftrightarrow\ket{1}$. Then, the
$2^{n-1}$ product vectors derived from (\ref{BellIneqOdd}) and
(\ref{BellIneqEven}) can be written as
\begin{eqnarray}\label{vectorsOdd}
V_{i_1}\ldots
V_{i_k}\sigma_{i_{1}-1}\ldots\sigma_{i_{k}-1}\ket{0}^{\ot
n},\nonumber\\
i_{1}<\ldots<i_{k}=1,\ldots,n,\quad
k=0,2,4,\ldots,n-1,
\end{eqnarray}
and
\begin{eqnarray}\label{vectorsEven}
V_{i_1}\ldots
V_{i_k}\sigma_{i_{1}-1}\ldots\sigma_{i_{k}-1}\ket{0}^{\ot
n},\nonumber\\
V_1V_{i_1}\ldots V_{i_k}\sigma_{i_{1}-1}\ldots\sigma_{i_{k}-1}\sigma_n\ket{0}^{\ot
n},\nonumber\\
i_{1}<\ldots<i_{k}=2,\ldots,n,\quad k=0,2,4,\ldots,n-2,
\end{eqnarray}
respectively. For $n=3$ we recover the
four-element three-qubit Shifts UPB 
\cite{BennettUPBPRL}. Notice that the freedom in choosing the
local sets allows one to obtain more general UPBs.

We are now ready to prove the following statement.

\noindent {\bf Fact 3.} {\it The vectors (\ref{vectorsOdd}) and
(\ref{vectorsEven}), form an $n$-qubit UPB.}

\noindent{\bf Proof.} Our proof consists of two steps. First we
show that for any $n$, the above vectors can be generated from the
Shifts UPB by a recursive protocol. Then we prove that this
protocol preserves the property of being UPB.

Let us start with the case of odd $n$. We denote the set of
vectors in Eq. (\ref{vectorsOdd}) by $U_1$ and divide it into two
subsets $U_{1}^{(i)}$ $(i=1,2)$ consisting of vectors with the
first qubit from $\mathcal{S}_{i-1}$. Then, we create another
group of vectors $U_2$ by switching the last qubit of $U_1$ to the
orthogonal one (henceforth called {\it orthogonalization}), and
divide $U_2$ into two subsets $U_2^{(i)}$ $(i=1,2)$ in the same
way as $U_1$. Finally, direct 
algebra shows that the following set of vectors
\begin{eqnarray}\label{UPBn1}
\ket{0}\ot U_1^{(1)}, \quad \ket{1}\ot U_{2}^{(2)}, \quad \ket{e}\ot U_{2}^{(1)},
\quad \ket{e^{\perp}}\ot U_1^{(2)},
\end{eqnarray}
is exactly the same as the vectors in (\ref{vectorsEven}) with
$n+1$ parties.

Almost exactly the same procedure produces $(n+1)$-partite vectors
(\ref{vectorsOdd}) from $n$-partite set with even $n$. The only
difference is that to obtain $U_2$ from $U_1$ one has to
orthogonalize the penultimate qubit and apply the transformation
$\ket{0}\leftrightarrow\ket{e^{\perp}}$ and
$\ket{1}\leftrightarrow\ket{e}$ to the last one.

Having established the recursive procedure generating vectors
(\ref{vectorsOdd}) and (\ref{vectorsEven}) from the Shifts UPB, we
now show that it preserves the UPB property. First, let us prove
that all the vectors (\ref{UPBn1}) are orthogonal. It suffices to
prove that vectors from $U_2$ are orthogonal, $U_1^{(1)}\perp
U_{2}^{(1)}$, and $U_1^{(2)}\perp U_{2}^{(2)}$ (notice that
already $U_{1}^{(1)}\perp U_1^{(2)}$). The first condition is
satisfied due to the fact that $U_2$ is obtained from $U_1$ by
application of the above local transformations. A direct check shows
that they map a set of orthogonal vectors onto another set of
orthogonal vectors.

The proof of the remaining two conditions is more involved.
Nevertheless, it suffices to consider the odd-$n$ case, since the
proof for the even $n$ goes along almost the same lines. To this
end, notice that the last qubit of the vectors in $U_1^{(1)}$ is
either $\ket{0}$ or $\ket{e}$ (cf. 
(\ref{UPBn1})). Thus, their
orthogonality comes from the first $n-1$ qubits. This, together
with the fact that $U_2^{(1)}$ is obtained from $U_{1}^{(1)}$ by
orthogonalizing the last qubit, imply that any vector from
$U_{1}^{(1)}$ is orthogonal to $U_{2}^{(1)}$ and hence
$U_1^{(1)}\perp U_{2}^{(1)}$. Exactly the same reasoning allows
one to conclude that $U_1^{(2)}\perp U_{2}^{(2)}$. The only
difference is that the last qubit of $U_1^{(2)}$ is either
$\ket{1}$ or $\ket{e^{\perp}}$ (also not orthogonal).

Finally, we show that there does not exist any product vector orthogonal to the
set (\ref{UPBn1}). For this purpose, assume the contrary and write the vector
orthogonal to (\ref{UPBn1}) as $\ket{\psi}=\ket{x}\ket{\widetilde{\psi}}$ with
$\ket{x}$ and $\ket{\widetilde{\psi}}$ denoting the first qubit and the product
state of the remaining $n-1$ qubits, respectively. If $\ket{x}$ belongs to one
of the sets $\mathcal{S}_{i}$ $(i=0,1)$, say $\mathcal{S}_0$, then
$\ket{\widetilde{\psi}}$ has to be orthogonal to either $U_1$ or $U_2$,
depending on whether $\ket{x}=\ket{0}$ or $\ket{x}=\ket{1}$. If $\ket{x}\notin
\mathcal{S}_i$ $(i=0,1)$, then $\ket{\widetilde{\psi}}$ must be orthogonal to
both UPBs $U_i$. Both situations lead to a contradiction meaning that the above
construction preserves the UPB property. $\blacksquare$

{\it Conclusions.---}Nontrivial Bell inequalities lacking a
quantum violation are rare and intriguing objects meriting further
investigation. We have demonstrated here a systematic way to
derive inequalities of this type with the property
$\beta_{\mathrm{C}}=\beta_{\mathrm{Q}}<\beta_{\mathrm{N}}$ from
UPBs, themselves an important concept in the theory of
entanglement. We have furthermore shown that the construction may
be applied in the reverse direction and have provided new examples
of UPBs from existing Bell inequalities.

These findings are strongly related to recent work 
on the relationship between QC and multipartite versions of
Gleason's theorem~\cite{Barnum10PRL,Hadley10PRL}.  Indeed, the
generalisation of this theorem to the case of distant observers
leads to correlations that can be written as local measurements
acting on entanglement witnesses. This set is equivalent to the
set of QC in the bipartite case~\cite{Barnum10PRL,Hadley10PRL}.
However, this equivalence does not hold for three
parties~\cite{Hadley10PRL}. Here, we generalize this result to any
Bell scenario in which one is able to build, by using our
procedure, a non-trivial inequality from some UPB satisfying (P).
On the other hand, our results imply that there are no bipartite
UPBs with the property (P), as otherwise there would exist a
bipartite witness violating a Bell inequality beyond the quantum
bound, contradicting the results of
\cite{Barnum10PRL,Hadley10PRL}.

The connection between product vectors and Bell inequalities
introduced here opens new perspectives. For instance, it is worth
investigating whether sets of orthogonal product vectors with (P),
which are not UPBs can lead to novel Bell inequalities. Although
all of them lack a quantum violation, it is unclear whether they
are nontrivial, \ie, violated by some nonsignalling correlations.
In this direction, we prove the following:

\noindent{\bf Fact 4.} {\it Let $\mathcal{S}$ be a completable set
of orthogonal product vectors with the property (P). Then the
corresponding Bell inequality (\ref{genBell}) is not violated by
any NC represented by entanglement witnesses.}

\noindent{\bf Proof.} Let $\mathcal{S}^{\perp}$ denote the set of
product vectors completing $\mathcal{S}$ to the full basis in
$\mathcal{H}$. Then consider the Bell inequality derived from
$\mathcal{S}$ and the Bell operator $B$ representing it (cf. 
[\ref{BellOp})]. By $\Pi$ let us now denote the separable
projector onto the support of $B$. The latter can act on a Hilbert
space of dimension larger than $\mathrm{dim}\mathcal{H}$, but,
since $\mathcal{S}$ is completable, there exists a separable
projector $\Pi^{\perp}$ such that $\Pi+\Pi^{\perp}=\mathbbm{1}$.
Then, for any normalized witness $\Tr(BW)\leq\max\{q_i\}\Tr(\Pi W)
=\max\{q_i\}\Tr[(\mathbbm{1}-\Pi^{\perp})W]$. As $\Pi^{\perp}$ is
separable and $W$ is normalized, $0\leq \Tr(W\Pi^{\perp})$ and
hence $\Tr(BW)\leq \max\{q_i\}$. $\blacksquare$

Finally it would be of interest to understand when this
construction leads to tight Bell inequalities, and 
if the ability to do so may be inferred from some properties of
the set of product states.  From a more general perspective, it
remains an open question as to whether there exist bipartite Bell
inequalities without a quantum violation.

{\it Acknowledgements.} We thank M. Piani, G. Prettico, and A.
Winter for discussions. This work was supported by EU projects
AQUTE, NAMEQUAM, Q-Essence, and QCS, ERC Grants QUAGATUA and
PERCENT, Spanish MINCIN projects FIS2010-14830, FIS2008-00784,
FIS2008-01236, FPU AP2008-03043, and QOIT, Generalitat de
Catalunya, and Caixa Manresa.

\end{document}